\documentclass[twocolumn,showpacs,preprintnumbers,amsmath,amssymb,widetext,super
scriptaddress]{revtex4}
\usepackage{graphicx}
\begin{document}
\title{Tailoring Josephson coupling through superconductivity-induced nonequilibrium}
\author{F. Giazotto}
\email{giazotto@sns.it} \affiliation{NEST-INFM \& Scuola Normale
Superiore, I-56126 Pisa, Italy}
\author{T. T. Heikkil\"{a}}
\affiliation{Low Temperature Laboratory, Helsinki University of Technology, P.O.
Box 2200, FIN-02015 HUT, Finland}
\author{F. Taddei}
\affiliation{NEST-INFM \& Scuola Normale Superiore, I-56126 Pisa,
Italy}
\author{Rosario Fazio}
\affiliation{NEST-INFM \& Scuola Normale Superiore, I-56126 Pisa,
Italy}
\author{J. P. Pekola}
\affiliation{Low Temperature Laboratory, Helsinki University of Technology, P.O.
Box 2200, FIN-02015 HUT, Finland}
\author{F. Beltram}
\affiliation{NEST-INFM \& Scuola Normale Superiore, I-56126 Pisa,
Italy}

\begin{abstract}
The distinctive quasiparticle  distribution  existing under
nonequilibrium in a superconductor-insulator-normal
metal-insulator-superconductor (SINIS) mesoscopic line is proposed
as a novel tool to  control  the supercurrent intensity in a long
Josephson weak link. We present a description of this system in
the framework of the diffusive-limit quasiclassical Green-function
theory and take into account the effects of inelastic scattering
with arbitrary strength. Supercurrent enhancement and suppression,
including a marked transition to a $\pi$-junction are striking
features leading to a fully tunable structure.
\end{abstract}


\maketitle

Nonequilibrium effects in mesoscopic  superconducting circuits have
been receiving a rekindled attention during the  last few years
\cite{articles}. The art of controlling Josephson coupling in
superconductor-normal metal-superconductor (SNS) weak links is at
present  in the spotlight: a recent breakthrough in mesoscopic
superconductivity is indeed represented by the  SNS transistor,
where supercurrent suppression  as well as its sign reversal
($\pi$-transition) were  demonstrated \cite{baselmans,baselmansphase}. This was
achieved by driving the quasiparticle distribution in the weak
link far from equilibrium \cite{volkov,wilhelm,yip} through external
voltage terminals, \emph{viz.} normal reservoirs. Such a  behavior
relies on
 the two-step shape of the quasiparticle nonequilibrium
 distribution, typical of diffusive mesoscopic wires and
experimentally observed by Pothier and coworkers \cite{pothier}.

The purpose of this paper is to demonstrate that  it is possible
 to tailor the quasiparticle distribution through
\emph{superconductivity-induced} nonequilibrium in order to
implement a unique class of superconducting transistors. This can
be achieved
 when mesoscopic control lines are connected to
superconducting reservoirs through tunnel barriers (I),
realizing a SINIS channel. The peculiar  quasiparticle
 distribution  in the N region, originating  from
 biasing the S terminals, allows one to access several regimes,  from supercurrent
enhancement with respect to equilibrium   to a large amplitude of
the $\pi$-transition passing through a steep supercurrent
suppression. These features are accompanied by a large current
gain (up to some $10^{5}$ in the  region of larger input impedance)
and reduced dissipation. The ultimate operating frequencies
available  open the way to the exploitation of this scheme for the implementation of ultrafast current amplifiers.

\begin{figure}[ht!]
\begin{center}
\includegraphics[width=8.5cm,clip]{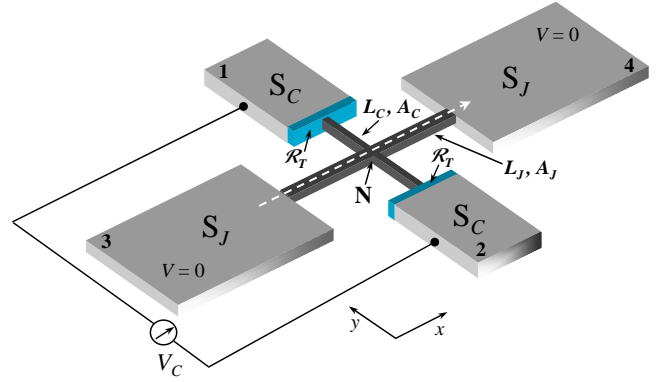}
\end{center}
\caption{Scheme of the Josephson transistor. The supercurrent
$I_J$ (along the white dashed line) is tuned by applying a bias
$V_C$ across the SINIS symmetric  line connected to the center of
the weak link. All normal wires are assumed
quasi-one-dimensional.} \label{transistor}
\end{figure}
The investigated mesoscopic structure  (see Fig.~\ref{transistor})
consists of a long diffusive weak link of length $L_J$ much larger
than the superconducting coherence length ($\xi_0$) oriented along
the $x$ direction. This defines the SNS junction of cross-section
$A_J$. The superconducting terminals belonging to the SNS
junction, labelled S$_J$ (3 and 4), are kept at zero potential.
The SINIS control line is oriented along the $y$ direction and
consists  of a normal wire, of length $L_C$ and cross-section
$A_C$, connected through identical tunnel junctions of resistance
$\mathcal{R}_T$ to two superconducting reservoirs S$_C$ (1 and 2),
biased at opposite voltages $\pm V_C /2$. The superconducting gaps
of S$_J$ and S$_C$ ($\Delta _J$ and $\Delta _C$) are in general
different.

The supercurrent $I_J$ flowing across the SNS junction is  given by
\cite{wilhelm,yip}
\begin{equation}
I_J(V_C)=\frac{\sigma A_J}{eL_J}\int_0^\infty dE\,
[f(-E;V_C)-f(E;V_C)]\textrm{Im}[j_E], \label{supercurrent}
\end{equation}
and depends on the quasiparticle distribution function $f(E)$. In
Eq.~(\ref{supercurrent}), $\sigma$ is the normal-state
conductivity which determines the normal-state resistance of the
junction according to $R_N =L_J /\sigma A_J$. The distribution
function $f$ reduces to the equilibrium Fermi distribution when
 $V_C =0$.  The energy-dependent spectral supercurrent
\cite{belzig,tero}, ${\rm Im}[j_E]$, can be calculated by solving
the Usadel equations \cite{usadel}. Following the parametrization
of the Green functions given in Ref. \cite{belzig}, these equations in the N
region can be written
\begin{equation}
\begin{split}
j_E =-\sinh ^2(\theta)\partial_x\chi,\,\,\,\,\,\,\,\,\,\,\,
\partial_xj_E=0, \\
\hbar D\partial_x
^2\theta+2iE\sinh\theta+\frac{\hbar D}{2}(\partial_x\chi)^2\sinh(2\theta)=0,
 \label{Usadel}
\end{split}
\end{equation}
where $D$ is the diffusion coefficient and $E$ is the energy
relative to the chemical potential in S$_J$. $\theta(x,E)$ and
$\chi(x,E)$ are in general complex functions. For perfectly
transmissive contacts, the boundary conditions at the S$_J$N
interfaces reduce to $\theta=\,$arctanh$(\Delta_J/E)$ and
$\chi=\pm\phi/2$ in the reservoirs $S_J$, where $\phi$ is the
phase difference between the superconductors.

\begin{figure}[ht!]
\begin{center}
\includegraphics[width=8.7cm,clip]{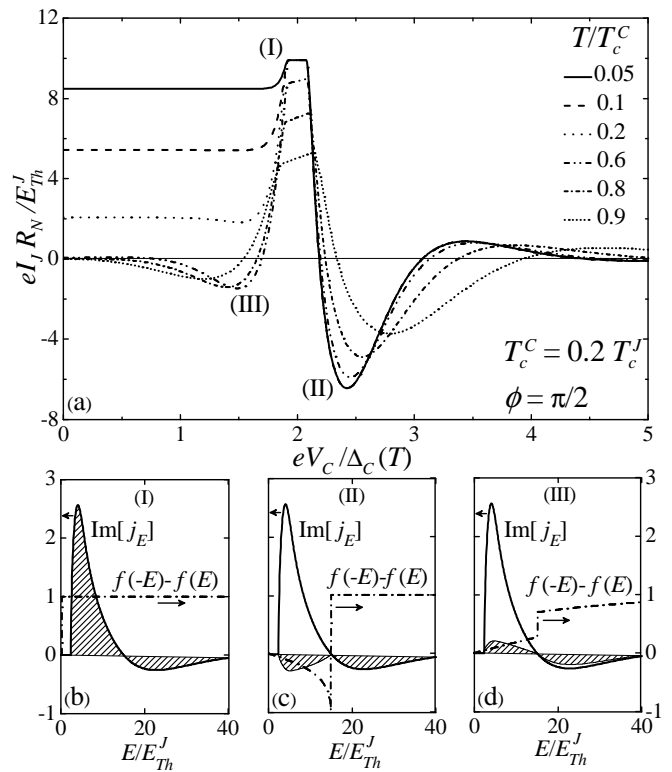}
\end{center}
\caption{(a) Supercurrent  vs control voltage $V_C$ at different
temperatures ($T$) for $\phi =\pi /2$ and $T_{c}^{C}=0.2\,T_{c}^{J}$ (see text).
Bias regions labeled (I), (II) and (III) indicate
supercurrent enhancement due to quasiparticle \emph{cooling},
high-voltage $\pi$-state and low-voltage $\pi$-state in the
high-temperature regime, respectively. These are qualitatively
explained in (b), (c) and (d) where hatched areas represent the
contribution to supercurrent arising in such bias ranges (see
text). } \label{supercurrentbehavior}
\end{figure}

As required by Eq.~(\ref{supercurrent}) we must determine the
actual quasiparticle distribution in the N region of the SINIS
structure. This is controlled by  voltage ($V_C$) and temperature,
and  by the amount of inelastic scattering in the control line. In
the case of a short control wire with no inelastic interactions,
the quasiparticle distribution, according to Ref.~\cite{Heslinga},
is given by
\begin{equation}
f(E,V_C)=\frac{\mathcal{N}_1 \mathcal{F}_1 +\mathcal{N}_2
\mathcal{F}_2}{\mathcal{N}_1 +\mathcal{N}_2},
 \label{noneq}
\end{equation}
where $\mathcal{N}_{1,2}=\mathcal{N}_{S_C}(E\pm eV_C/2)$ and
$\mathcal{F}_{1,2}=\mathcal{F}^0(E\pm eV_C/2 )$. The former are
the BCS densities of states in the reservoirs S$_C$ (labeled by 1
and 2 in Fig.~\ref{transistor}). $\mathcal{F}^0(E)$ is the Fermi
function at lattice temperature $T$ \cite{lowenote}. In this case
Eqs. (\ref{supercurrent}) and (\ref{noneq}) yield the
dimensionless transistor output characteristics shown in Fig.
\ref{supercurrentbehavior}(a). The latter plots the supercurrent
$I_J$ versus control bias $V_C$ at different temperatures for a
long junction (i.e., $\Delta_J \gg E_{Th}^J$, where $E_{Th}^J
=\hbar D/L_J ^2$ is the Thouless energy of the SNS junction, as
this is the limit where the supercurrent spectrum varies strongly
with energy). We assumed $\phi=\pi/2$, $T_c ^C /T_c ^J =0.2$,
where $T_c ^{C(J)}$ are the critical temperatures of the
superconductors S$_{C(J)}$ and $L_J$ such that $\Delta_J /E_{Th}^J
=300$.

At the lowest temperatures, by increasing $V_C$ curves display
a large supercurrent enhancement with respect to equilibrium
at bias $V_C = 2\,\Delta_C (T)/e=V_C ^\ast (T)$ (region I in
the figure). Further increase of bias leads to a
$\pi$-transition (region II) and finally to a decay
for larger voltages. This behavior is explained in
Figs.~2(b,c,d) where the spectral supercurrent (solid line) is
plotted together with $f(-E)-f(E)$ (dash-dotted line) for values
of $V_C$ and $T$ corresponding to regions I, II and III,
respectively. Hatched areas represent the integral of their
product, i.e., the supercurrent $I_J$ of Eq. (\ref{supercurrent}).
In particular, region I corresponds to the \emph{cooling} regime
where hot quasiparticles are extracted from the normal metal
\cite{Heslinga,leivo}. The origin of the $\pi$-transition in region II
is illustrated by Fig.~\ref{supercurrentbehavior}(c), where the
negative contribution to the integral is shown.  We remark
that the intensity of the supercurrent inversion is very
significant. It reaches about 60$\%$ of the maximum value of $I_J$
at $V_C \simeq V_C ^\ast (T)$ in the whole temperature range,
nearly doubling the $\pi$-state value of the supercurrent as
compared to the case of an all-normal control channel
\cite{wilhelm,yip}.
 In the high-temperature regime ($T/T_c^C\gtrsim 0.6$), when the equilibrium
critical current is vanishing, the supercurrent first undergoes a
low-bias $\pi$-transition (region III in the figure), then enters
regions I and II. This recover of the supercurrent
from vanishingly small values at equilibrium
is again the consequence of the peculiar shape of $f$ (see
Fig.~\ref{supercurrentbehavior}(d)). Notably, the supercurrent
enhancement around $V_C ^\ast(T)$ remains  pronounced even at the
highest temperatures, so that $I_J$ attains values largely
exceeding 50$\%$ of the junction maximum supercurrent.
 This demonstrates the
full tunability  of the supercurrent through nonequilibrium
effects induced by the superconducting control lines. We remark that
this is a unique feature stemming from the superconductivity-induced
nonequilibrium population in the weak link.

The length $L_C$ of the SINIS control line can be additionally
varied to control the supercurrent by changing the effective
strength of inelastic scattering in the N region. For
$\mathcal{R}_T \gg R_C$, the distribution function $f(E)$ in the N
region is essentially $y$-independent and we have
\begin{equation}
\begin{split}
&\frac{1}{e^2 \mathcal{R}_T \Omega_C
\nu_F}[\mathcal{N}_1(\mathcal{F}_1-f(E))+
\mathcal{N}_2(\mathcal{F}_2-f(E))]\\&+\kappa \int d\omega
d\varepsilon \,\omega^{\alpha}
\mathcal{I}(\omega,\varepsilon,y,E)=0.
\end{split}
\label{kinetic}
\end{equation}
Here $\nu_F$ is the normal-metal density of states at the Fermi
energy, $\Omega_C$ is the volume of the N region and $\mathcal{I}$
is the net collision rate at energy $E$. At low temperatures, the
most relevant scattering mechanism is electron-electron scattering
\cite{anthore} and  we can neglect the effect of electron-phonon
scattering. Then \cite{nag,pothier},
\begin{equation}
\mathcal{I}(\omega,\varepsilon,y,E)=\mathcal{I}^{in}(\omega,\varepsilon,y,E)-\mathcal{I}^{out}(\omega,\varepsilon,y,E),
\label{collision1}
\end{equation}
and
\begin{eqnarray}
\mathcal{I}^{in}(\omega,\varepsilon,E)=(1-f({\varepsilon}))(1-f(E))f({\varepsilon-\omega}) f(E+\omega),\\
\mathcal{I}^{out}(\omega,\varepsilon,E)=(1-f({\varepsilon-\omega}))(1-f({E+\omega}))f({\varepsilon})f(E).
\label{collision2}
\end{eqnarray}
Electron-electron interaction is either due to
direct Coulomb scattering \cite{AA,kamenev} or mediated by
magnetic impurities \cite{anthore}. Below, we concentrate on the
former, but the latter would yield  a similar qualitative
behavior. From the calculation of the screened Coulomb interaction
in the diffusive channel, it follows \cite{AA} that $\alpha =-3/2$
for a quasi-one dimensional wire and $\kappa
=(\pi\sqrt{D/2}\hbar^{3/2}\nu_{F}A_C)^{-1}$ \cite{kamenev}. We
note that $\Delta_C$ is the most relevant energy scale  to
describe  the distribution function for different voltages $V_C$.
It is thus useful to replace $\omega\rightarrow \omega/\Delta_C$
and $\varepsilon\rightarrow \varepsilon/\Delta_C$, in order to
obtain a dimensionless equation. Multiplying Eq.~(\ref{kinetic})
by $e^2 \mathcal{R}_T \Omega_C \nu_F$ we obtain
\begin{equation}
\begin{split}
\mathcal{N}_1(\mathcal{F}_1 -f(E))-
\mathcal{N}_2(f(E)-\mathcal{F}_2)=\\
=\mathcal{K}_{coll} \int d\omega d\varepsilon \omega^{-3/2}
\mathcal{I}_{coll}^{tot}(\omega,\varepsilon,y,E),
\end{split}
\label{SINISdistribution}
\end{equation}
where
$\mathcal{K}_{coll}=\frac{\mathcal{R}_T}{R_C}\frac{L_C^2\kappa}{D}\sqrt{\Delta_C}=\sqrt{2}\frac{\mathcal{R}_T}{R_K}\sqrt{\frac{\Delta_C}{E_{Th}^C}}$,
$R_K =h/2e^2$ and $E_{Th}^C =\hbar D/L_C ^2$. In the absence of
electron-electron interaction ($\mathcal{K}_{coll}=0$) Eq.
(\ref{noneq}) is  recovered.
\begin{figure}[h!]
\begin{center}
\includegraphics[width=8.7cm,clip]{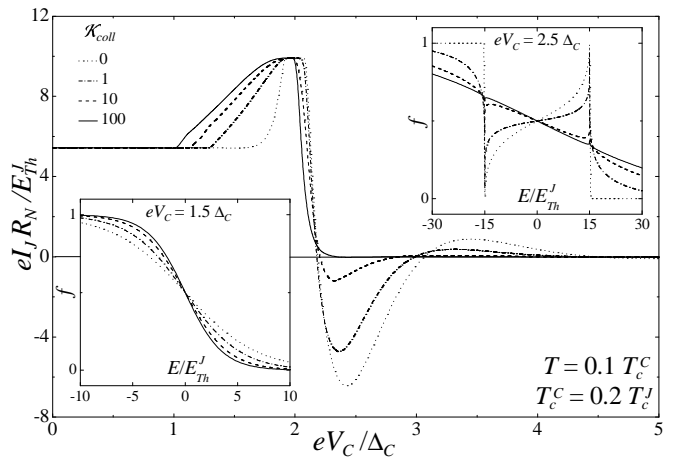}
\end{center}
\caption{Supercurrent  vs  $V_C$ for various $\mathcal{K}_{coll}$
with $T=0.1\,T_c ^C$ and $T_{c}^{C}=0.2\,T_{c}^{J}$.
Insets show the distribution function at
 $eV_C =1.5\Delta_C$ (left) and  $eV_C
=2.5\Delta_C$  (right) calculated for the same
$\mathcal{K}_{coll}$ values.} \label{SINISdistrfig}
\end{figure}

The influence of inelastic scattering on $I_J$ is shown in
Fig.~\ref{SINISdistrfig}, which displays the critical current of a
long junction  at $T=0.1\,T_c ^C$ for several values of
$\mathcal{K}_{coll}$. Here $I_J$ is obtained by numerically
solving Eq. (\ref{SINISdistribution}). The effect of
electron-electron interaction is to strongly suppress the
$\pi$-state and to widen the
 peak around $V_C ^\ast$. The $\pi$-transition
 vanishes for $\mathcal{K}_{coll}\simeq 100$, but
the $I_J$ enhancement due to quasiparticle cooling still persists
in the limit of even larger inelastic scattering \cite{giazotto}.
The disappearance of the $\pi$-state can be understood by looking
at the right inset of Fig.~\ref{SINISdistrfig} which clearly shows
how $f$ (calculated at $eV_C =2.5\Delta_C$) gradually relaxes from
nonequilibrium towards a Fermi  function upon increasing
$\mathcal{K}_{coll}$. The left inset shows how  $f$ (evaluated at
$eV_C =1.5\Delta_C$) sharpens, thus enhancing $I_J$, by increasing
$\mathcal{K}_{coll}$. This effect follows from the fact that
inelastic interactions redistribute the  occupation of
quasiparticle levels in the N region, thus increasing the
occupation
 at higher energy. As a consequence, higher-energy
excitations are more effectively
 removed  by tunneling, even for biases well below and
not only around $V_C ^\ast$ (as in the case of
$\mathcal{K}_{coll}=0$). At the same time,
supercurrent recovery at high temperature is gradually weakened upon
enhancing $\mathcal{K}_{coll}$. Notably, these calculations show
that a rather large amount of inelastic scattering is necessary to
weaken and completely suppress the $\pi$-state. For example, using
Al/Al$_2$O$_3$/Cu as materials composing the SINIS line,
$\mathcal{K}_{coll}= 1$ corresponds to use a fairly long control
line with $L_C\simeq2.3$ $\mu$m \cite{device}.

\begin{figure}[h!]
\begin{center}
\includegraphics[width=8.7cm,clip]{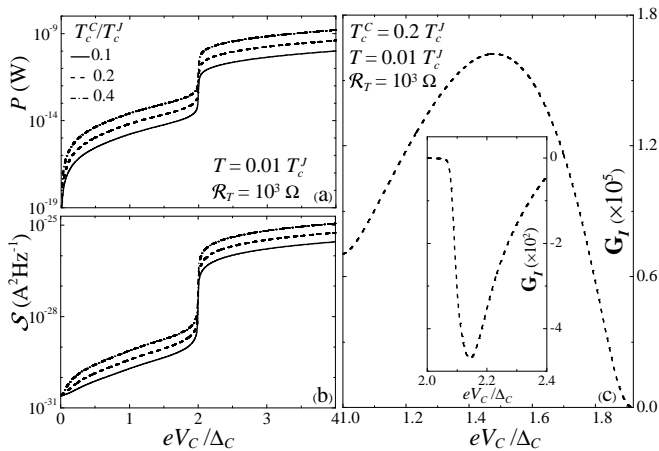}
\end{center}
\caption{(a) Power dissipated in the SINIS line vs $V_C$
calculated for various ratios $T_c ^C /T_c ^J$ and $T =0.01 T_c
^J$. (b) Noise power $\mathcal{S}$ vs $V_C$ calculated for the
same parameters as in (a). (c) Differential current gain
$\mathbf{G}_I$ vs $V_C$ for $T_c ^C /T_c ^J =0.2$. The inset shows
$\mathbf{G}_I$ in the high-bias region. In all these calculations
we set $\mathcal{K}_{coll}=0$ and $T_c ^J =9.26$ K (Nb).} \label{transistorchar}
\end{figure}

Changing the  ratio $T_c ^C /T_c ^J$
   shifts the $I_J$ response along the $V_C$ axis, the shape
of the characteristics being virtually independent of $T_c^C$.
This translates into a different magnitude of control voltages
$V_C$ and  power dissipation $P=I_C V_C$, where $I_C$ is the
control current across the SINIS channel. The function $P(V_C)$ is
plotted in Fig.~\ref{transistorchar}(a) for some ratios $T_c ^C
/T_c ^J$ at $T=0.01 \,T_c ^J$, assuming $\mathcal{R}_T =10^3
\,\Omega$ and $T_c ^J =9.26$ K. The impact of $\Delta_C$ in
controlling power dissipation is easily recognized. These effects
clearly indicate that $\Delta _C \ll \Delta _J$ is the condition
to be fulfilled in order to minimize $P$. In practice, the power
dissipation for $V_C > V_C ^\ast$ constitutes an experimental
problem as this energy needs to be carried out from the
reservoirs. In a similar way the noise properties of the system
are sensitive to the different $T_c ^C /T_c ^J$ ratios.
Assuming that the noise through one junction is essentially
uncorrelated from the noise through the other, it follows that the
input noise power $\mathcal{S}$ in the control line can be
expressed as
\begin{equation}
\mathcal{S}=\frac{1}{\mathcal{R}_T}\int
_{-\infty}^{\infty}dE\mathcal{N}_1
[f(E)(1-\mathcal{F}_1)+\mathcal{F}_1(1-f(E))]. \label{noise}
\end{equation}
$\mathcal{S}(V_C)$ from Eq. (\ref{noise}) is shown in
Fig.~\ref{transistorchar}(b) for the same parameters of
Fig.~\ref{transistorchar}(a). For example, for $T_c ^C /T_c ^J
=0.1$ (corresponding roughly to the combination Al/Nb), $P$
obtains values of the order of a few  $10^{-15}$ W and
$\mathcal{S}$ of some $10^{-30}$ A$^2$Hz$^{-1}$ in the cooling
regime, while these values are enhanced respectively to  few tens
of $10^{-12}$ W and $10^{-26}$ A$^2$Hz$^{-1}$ for biases around
the $\pi$-transition.

In light of the possible use of this operational principle for
device implementation, let us comment on the available gain and
switching times. Input  and output
 voltages are of the order of $\Delta _C /e$
and $E_{Th}^J /e$, respectively, so that it seems hard to achieve
voltage gain. On the other hand, differential current gain
$\mathbf{G}_I =dI_J /dI_C$ can be very large. For $V_C
>V_C ^\ast$ a simple estimate  gives
$\mathbf{G}_I \sim (E_{Th}^J/\Delta _C)(\mathcal{R}_T /R_N)$,
meaning that with realistic ratios $\mathcal{R}_T /R_N$ ($\sim
10^3$), $\mathbf{G}_I$ can
 exceed $10^2$.
$\mathbf{G}_I (V_C)$ calculated for $T_c ^C /T_c ^J =0.2$ is
plotted in Fig.~\ref{transistorchar}(c) (the inset shows the gain
in the $\pi$-state region). This calculation reveals that
$\mathbf{G}_I$ can reach huge values, with some $10^{5}$ for $V_C
<V_C ^\ast$ \cite{gain} and several $10^{2}$ in the opposite
regime. Remarkably, gain is almost unchanged also in the presence
of weak inelastic scattering (i.e., $\mathcal{K}_{coll}=1$). The
same holds for $P$ and $\mathcal{S}$. The highest operating
frequency $\nu$ of the transistor is limited by the smallest
energy in the system: $\nu \leq \min \frac{1}{h}\{\Delta_C
,\Delta_J ,E_{Th}^C ,E_{Th}^J , h(\mathcal{R}_T
\mathcal{C})^{-1}\}$ where $\mathcal{C}$ is  the tunnel junction
capacitance. For an optimized device, working frequencies of the
order of $10^{11}$ Hz can be experimentally achieved in the
high-voltage regime $V_C > V_C ^\ast$. For $V_C <V_C ^\ast$,
conversely, the response is slower (somewhat below $10^9$ Hz),
owing  to the long discharging time through the junctions.

We thank M. H. Devoret, K. K. Likharev, F. Pierre, L. Roschier, A.
M. Savin, and V. Semenov for helpful discussions. This work was
supported in part by MIUR under the FIRB project RBNE01FSWY and by
the EU (RTN-Nanoscale Dynamics).



\begin{thebibliography}{25}
\bibitem{lowenote}
    At low control voltages there is a region of energies where
$\mathcal{N}_1=\mathcal{N}_2=0$. We assume that there the
(otherwise weak) coupling to phonons makes the distributions at
those energies equal to the equilibrium Fermi distribution. A more
detailed discussion and another type of coupling is given in
\cite{cooler}.
\bibitem{device}
     This is straightforward assuming as typical parameters  $\mathcal{R}_T=10^3$ $\Omega$,
    $D=0.02$ m$^2$/s and $\Delta_C =200$ $\mu$eV.
\bibitem{gain}
    In this calculation we chose to include depairing by a phenomenological, but realistic parameter $\Gamma = 10^{-4} \Delta_C$ \cite{cooler}.
Its omission would lead to extremely higher $\mathbf{G}_I$
values.
\end{thebibliography}
\end{document}